**Title:** Leveraging the Global Research Infrastructure to Characterize the Impact of National Science Foundation Research

**Authors:**

- Jamaica Jones, https://orcid.org/0000-0002-1969-2508, University of Pittsburgh (https://ror.org/01an3r305), School of Computing and Information, 135 North Bellefield Avenue, Pittsburgh, PA 15213. jaj160@pitt.edu
- Ted Habermann, https://orcid.org/0000-0003-3585-6733, Metadata Game Changers (https://ror.org/05bp8ka05)

**Abstract:**

The Global Research infrastructure (GRI) is made up of the repositories and organizations that provide persistent identifiers (PIDs) and metadata for many kinds of research objects and connect these objects to funders, research institutions, researchers, and one another using PIDs. The INFORMATE Project has combined three data sources to focus on understanding how the global research infrastructure might help the US National Science Foundation (NSF) and other federal agencies identify and characterize the impact of their support. In this paper we present INFORMATE observations of three data systems. The NSF Award database represents NSF funding while the NSF Public Access Repository (PAR) and CHORUS, as a proxy for the GRI, represent two different view of results of that funding. We compare the first at the level of awards and the second two at the level of published research articles. Our findings demonstrate that CHORUS datasets include significantly more NSF awards and more related papers than does PAR. Our findings also suggest that time plays a significant role in the inclusion of award metadata across the sources analyzed. Data in those sources travel very different journeys, each presenting different obstacles to metadata completeness and suggesting necessary actions on the parts of authors and publishers to ensure that publication and funding metadata are captured. We discuss these actions, as well as implications our findings have for emergent technologies such as artificial intelligence and natural language processing.

# Introduction

The global research infrastructure (GRI) is made up of the repositories and organizations that provide persistent identifiers (PIDs) and metadata about many kinds of research objects (preprints, published papers, datasets, dissertations, proposals, reviews, etc.) and connect these objects to funders, research institutions, researchers, and one another using PIDs. This infrastructure currently contains millions of objects and is growing rapidly in every possible direction.

CHORUS brings together funders, societies, publishers, and institutions from across the open research ecosystem to share knowledge, develop solutions, advance innovation, and support collective efforts. CHORUS retrieves data from across the GRI and provides open services for users: a search, a dashboard, a public API, and a series of reports.[1] When combined with the API, these data serve as proxies for the GRI that can be used for data exploration and analysis that supports insight into open access and the impact of research funding across a variety of federal agencies.

*Informating*, a word coined by Zuboff, is the process that translates descriptions and measurements of activities, events and objects into information.[2] Drawing its name from this definition, the INFORMATE Project has combined three data sources to focus on understanding how the global research infrastructure might help the US National Science Foundation (NSF) and other federal agencies identify and characterize the impact of their support.[3] The questions we have focused on include:

- How can this infrastructure improve identification of the myriad contributions made to global knowledge by funders like the National Science Foundation and other federal agencies?
- How can we use this infrastructure to increase understanding of connections across the US and global research landscape?

- How can this infrastructure be used to increase completeness, consistency, and connectivity within agency repositories and search tools?

In this paper we present INFORMATE observations of three data systems. The NSF Award database represents NSF funding while the NSF Public Access Repository (PAR) and CHORUS, as a proxy for the GRI, represent two different views of results of that funding. We compare the first two at the level of awards and the second two at the level of published research articles, identified using Digital Object Identifiers (DOIs).[4,5] In so doing, we contribute to a growing body of research leveraging the global research infrastructure in bibliometric analyses of state-sponsored research outputs.

These studies often center DOIs and funder and award metadata in efforts to assess research impact. For example, Gerasimov et al. sought to characterize the scope of peer-reviewed articles citing 11,000 NASA earth science datasets registered with DOIs within the agency's Earth Observing System Data and Information System (EOSDIS). Drawing upon metadata from common bibliographic sources including Scopus, Web of Science, Crossref and Google Scholar, the team discovered over 17,000 "dataset-citing publications" and amid their results note that completeness across sources varies considerably.[6]

A later study by Eric Schares undertook to estimate the impact of the 2022 White House Office of Science and Technology Policy (OSTP) Public Access ("Nelson") Memo by finding connections between publications and funder metadata. In this study, Schares relied upon the global research product Dimensions to quantify the number of publications attributed to US federal funding and then extrapolate a total number of such publications that might be affected by Nelson Memo guidance.[7] The opening line of Schares' "Limitations" section acknowledges that the "clearest limitation of [his] analysis is the likelihood that not all U.S. federally funded research is included in the dataset" he used.[8] Attributing this constraint to the bounds of Dimensions' indexing and by extension to the journey traveled by funder metadata across the GRI, Schares traces the source of Dimensions' funder metadata to Crossref.

   Prior research by Nees van Eck and Ludo Waltman has suggested that Crossref's funder metadata is dependent on the publisher providing it. [9] This is corroborated by Bianca Kramer and Hans de Jonge, who examined a corpus of over 5,000 publications known to have resulted from the funding of the Dutch Research Council NKO.[10] Seeking to determine the completeness of Crossref funder metadata associated with these publications, the team found that, while coverage was on the whole strong, just over half of the publications analyzed referenced NWO by name and just under half included the funder ID.[10] Homing their focus on the ways in which three bibliographic databases (Web of Science, Scopus, and Dimensions) process information embedded in the acknowledgement sections of the publications analyzed, they observe that, "there are considerable differences in the extent to which the three… databases succeed in extracting funding information for publications of different publishers. This seems to suggest that [they] differ in the extent to which they have access to the content of various publishers."[10] Across relevant literature, then, it emerges that an understanding of metadata provenance, encompassing broad and deep insight into the journeys traveled by bibliometric metadata, is critical to accurately interpreting the results of analyses related to the global research infrastructure.

## Data Sources and Journeys

   Bates, Lin and Goodale introduce the concept of a data journey as a means of "illuminating the socio-material life of data as they travel between and through different sites of data practice."[11] Distinguishing these paths from the purportedly seamless flow of data described by others in the field, they argue that "journey" is a term more appropriate to the episodic movement of data through networked systems, noting the "disjointed breaks, pauses, start points [and] end points… that occur as data move, via different forms of

'transportation',... between different sites of data practice across knowledge infrastructures."[11]

Adopting this as a perspective through which one might examine the life of research metadata provides meaningful insight into how particular views into the GRI shape one's understanding of it. The three primary datasets we used have significantly different data journeys ranging from simple (NSF Awards) to complex (CHORUS). Our aim in this section is to describe these data journeys as they provide critical context for understanding and comparing data across the sources.

## NSF Award Search

The US National Science Foundation is a major contributor to the global research ecosystem, providing roughly a quarter of federal funding for basic research across US colleges and universities.[12] NSF Award Search is a public database that provides download and API access to roughly 70 years of records - including information regarding principal investigators, funded projects, and funding amounts. While many of the older records accessible through NSF Award Search lack some detail, those dating to 1976 and later provide a sweeping view into the agency's funding history.[13]

All metadata in NSF Award Search records are created and managed by the institution itself, resulting in a short data journey. Because these data are generated directly from the source, they offer an extensive accounting of NSF awards granted. For the purposes of this study, we have taken NSF Award Search data to represent the complete "known universe" of NSF award numbers.

## NSF Public Access Repository (PAR)

When principal investigators publish NSF-funded findings in peer-reviewed journals, they are required to submit a copy of each article to NSF's Public Access Repository as part of NSF award reporting procedures. While deposit of peer-reviewed articles is required, other types of research outputs are also accepted. These resources are described by bibliographic and other metadata documenting author, publication and award information, the majority of which is provided by authors at the time of PAR submission.

This workflow comprises the most common manifestation of the NSF PAR data journey. It is a short trip, dependent on the diligence of the person doing the reporting, the constraints of the reporting interface, and the contribution and interventions of NSF's internal data management systems. These include the synching of certain key metadata such as NSF Award IDs as well as the addition of other critical elements such as OSTI ID (a unique value identifying each record) and record entry date, and data improvement efforts made after record creation. Despite the short length of the PAR data journey, it varies on a case-by-case basis and introduces inherent, irreproducible elements. In addition, there are obstacles that prevent inclusion of funded works, so NSF PAR is an incomplete sample of the universe of works funded by NSF.

## CHORUS

CHORUS is an organization that works with funders, publishers and other institutions across the GRI to "share knowledge, develop solutions, advance innovation and support collective efforts".[1] CHORUS reports are populated with data drawn from across the GRI, connecting publications, datasets, authors, awards and other entities to funders such as NSF. The reports are created for several funders and are available from the CHORUS Dashboard (Figure 1).[14] The CHORUS reports used in this work provide an overview of metadata for journal articles (the All Report), authors (the Author Affiliation Report), and datasets (the Dataset Report). Together these three reports include 83 metadata elements. CHORUS includes significantly more articles than PAR (see details below) so it is currently our best estimate of the "known universe" of funded work.

In addition to access to reports, the CHORUS dashboard provides visualizations of histories of many open science measures (Figure 1). Three of these histories are directly related to the reports that we used: the Total curve (black) reflects the total number of articles which acknowledge funding from a specific funder, the ORCID IDs curve (pink) reflects the number of those articles that include one or more ORCIDs, and the Dataset curve (green) reflects the number of datasets connected to the articles.

The global infrastructure and its plumbing change regularly; understanding those changes and their impacts is critical for understanding the data and interpreting them correctly. The histories in Figure 1 for NSF (as well as in the CHORUS Dashboard for other funders) are generally smooth, but they have some large jumps, for example, the number of ORCIDs (pink in Figure 1) increased from ~145,000 to ~300,000 during March 2024. Similar jumps occurred at the same time for other funders. For example, USGS ORCIDs jumped from 2,655 to 4,628, and USAID ORCIDs jumped 2,839 to 5,614. The fact that these jumps occurred across many funders suggests that they are embedded in the data journeys and CHORUS' responses to those changes rather than being changes in the underlying content. They illustrate the need for clear understanding of those journeys before interpreting the numbers.

Of the three datasets we considered, CHORUS has the most complex data journey. The first step towards understanding these journeys is understanding the data sources. Figure 2 shows the three reports considered (green), the metadata they include, and the sources of the metadata. The All Report near the top of Figure 2 includes journal article metadata from Crossref (orange) and administrative metadata added by CHORUS (yellow). The Dataset Report on the left includes primarily metadata from DataCite (red) although there is some metadata from Crossref and, critically, dataset DOIs from ScholeXplorer (purple). The Author Affiliation report on the right includes primarily metadata from Crossref and ORCID. Some elements, near the center of the Figure, are shared by all three reports and some, in the upper right, are shared by the All Report and the Authors Report. The connections shown in Figure 2 illustrate that while these reports rely mostly on single sources, they are also connected to provide an integrated picture of articles, datasets, and researchers.

The data journey to CHORUS shown in Figure 3 starts with funding agencies (with Funder Identifiers) supporting researchers (with ORCIDs) who write journal articles and create datasets. The articles are submitted to and published in journals that share metadata in Crossref where digital object identifiers (DOI) are minted for articles. Some researchers also collect data and register it with institutional repositories and/or with DataCite to receive a dataset DOI.

Article and dataset identifiers and the metadata associated with them are the lifeblood of the infrastructure and of CHORUS. The researchers are the primary source for most of the metadata associated with these articles and datasets, but in both cases, intermediaries, either publishers or institutional repositories, may influence the metadata that makes it into the global research infrastructure.

Once articles are accepted and metadata are added to Crossref, CHORUS queries those metadata through the Crossref API using the Funder Name and the Funder Identifier (A in Figure 3) and retrieves metadata (B in Figure 3) for journal articles that acknowledge the funder(s) supporting their work.(15) This Crossref metadata forms the basis for all CHORUS reports (C in Figure 3, also Figure 2) and the Crossref DOIs are the keys to more detailed metadata for authors and datasets.

CHORUS queries ScholeXplorer, a collection of over 300 million links, with the article DOIs (D in Figure 3) to find datasets linked to the articles, and dataset DOIs for those datasets.[16] Those DOIs (purple in Figure 3) go into the All Report and are used to query the DataCite API to retrieve dataset metadata that goes into the CHORUS Dataset report along with some Crossref metadata (E in Figure 3; also Figure 2).[17]

Understanding this journey is important because it includes several obstacles that must be overcome for datasets to be included in the CHORUS Datasets Report:

- The researchers must provide funder metadata (name and identifier) for their article and the journal publisher must provide that metadata to Crossref,
- The dataset created or used must be in a repository and must have a DataCite DOI,
- The link between the article and the dataset DOI must be included in ScholeXplorer.

The journey to the Authors report starts with a query to the ORCID API using the article DOI from Crossref (F in Figure 3).[18] This query returns ORCIDs that are associated with the DOI that are then included in the CHORUS All and Author Reports along with metadata from the original Crossref query (G in Figure 3). The metadata in the Author report faces obstacles similar to those mentioned above:

- The researchers must have ORCIDs,
- The metadata associated with the ORCIDs must be publicly available,
- The ORCIDs must be associated with the articles.

Each step in this data journey includes obstacles that must be overcome for data to make it into CHORUS reports. Similar obstacles exist regardless of the mechanism used to retrieve data from the GRI and must be considered in analysis and interpretation of GRI results from any source. For example, it is not unusual to use Crossref metadata to find resources funded by particular funders but, to be found this way, the articles must include structured funder metadata (funder ID, award number). Figure 4 shows the percentage of journal articles in Crossref with funder metadata as a function of time. The coverage increases over time with a sharp jump so far in 2024, but it averages less than 25%. Our observations and interpretations must all be tempered by the fact that, because they start with a Crossref query, they currently only apply to a small slice of the publication universe which varies over time.

The paucity of funder metadata shown in Figure 4 emphasizes the critical role of the publishers at the beginning of the data journey and their interaction with the researchers (the primary metadata source, Figure 3). Eleven publishers are responsible for approximately 80% of the articles included in the CHORUS dataset so the impact of their practices is significant. In particular, the improvement of mechanisms for collecting and formatting funder metadata could improve the entire dataset and measures of funder impact derived from it.

## Methodology

The first step in our analysis was to collect data from their respective sources. Each data source required some data cleanup prior to analysis. Common steps are summarized in Table 1 with more details described below.

| Source | Origin | Temporal Coverage | Process Notes |
|---|---|---|---|
| NSF Awards | NSF | 2004 - 2021 | Yearly XML source parsed to CSV (Open Refine) or yearly CSV files.(19) Only continuing and standard grants and fellowship awards considered. Date strings converted to years |
| NSF PAR | PIs | 2015 – June 2022 | Yearly XML source parsed to CSV (Open Refine) Date strings converted to years |

| | | | DOIs reformatted when necessary. Well-formed award numbers (seven digits) |
|---|---|---|---|
| CHORUS | Crossref with Funder Metadata | 2014 - 2024 | All Report xlsx download from CHORUS – reformatted to CSV Publication Years created from earliest of Online and publication dates Award IDs separated from semi-colon separated values, invalid values are common. |

Table 1. Data clean-up applied to each data source.

NSF Award Search data were downloaded by year from the NSF Award Search website. Encoding inconsistencies in the source data required a pair of approaches in the preliminary data processing. Some yearly files were downloaded as XML and parsed into CSV using Open Refine. Others were downloaded directly as CSV. In both cases, data were limited to records for continuing grants, standard grants and fellowship awards, as these were presumed to be the most likely to result in published outputs. The relatively short data journey traveled by NSF Award Search data results in it being remarkably clean and standardized. As a result, very little cleaning was required.

NSF PAR data were received via email directly from NSF in 2022, prior to the start of this study. This dataset contains all PAR records that were active at that point in time and, as such, documents PAR through June 2022. PAR was established during 2015, so no records were created before that year. However, the database does contain records for publications published before 2015.

These data are organized by research output; each research output is assigned a unique identifier called the "result - osti_id". In the case of records with multiple authors or supporting awards, each new value is documented in a new row, resulting in a "record" that is distributed across many rows of data. Figure 5 illustrates this one-to-many relationship between publication and authors; the same relationship exists between publications and awards.

Iterative treatment was needed to address extensive errors in PAR DOI values. This involved a variety of programmatic approaches to correct for variations in DOI formatting. In the PAR dataset received, all original DOI values had been prepended with the prefix "https://doi.org/" regardless of the formatting of those original values. This resulted in some values containing two "https://doi.org/" prefixes. While this presented perhaps the single most common error in PAR DOI values, many other errors existed - a vastly heterogeneous mix of syntactical errors captured by the free text field in the NSF PAR grant reporting system. Figure 6 provides a snapshot depicting the variation in DOI value syntaxes contained in the original PAR data. The most common of these were identified and corrected using find and replace methodologies. Due to the volume of data processed during this study, the resulting DOI values were not validated.

CHORUS reports, requested for all publication outputs linked to NSF funding, were downloaded directly from CHORUS and subjected to a series of transformations similar to those used to normalize NSF Award and NSF PAR data. As is the case with the NSF PAR data, the CHORUS All report also manages one-to-many relationships between publications and authors, and publications, funders and award numbers. Since our analysis was completed, CHORUS has modified its approach to handling repeated elements. At the time of our analysis, repeated elements were concatenated into semicolon-separated values stored in a single field. In the case of award identifiers, for instance, all award numbers used to reference grant funding from any source were concatenated into a single "GRANT ID" field, some with prefixes suggesting the source of the funding (e.g.: NIH:R01 LM010730), and some without (e.g.: SGH16B008). Regular expressions were therefore necessary to identify all strings with the syntax of a well-constructed NSF Award ID. These were extracted and paired in new rows with the corresponding DOIs of publications that emerged from each award. These unique pairs of DOIs and NSF Award IDs therefore formed the unit of analysis used in our work with the CHORUS All report.

Comparisons across these data sources raise questions at the award and article-levels. Here, we will present our methodology and findings relevant to each, focusing the data sources relevant at each level.

## Articles

A primary research question guiding our work regards article occurrence and agreement across data sources. Specifically, comparisons between CHORUS and PAR data allow us to understand the extent to which NSF funded publications are represented across the global research infrastructure and within the institution's own repository; they likewise allow us to identify gaps in coverage.

Searching for content in PAR begins at the PAR interface, which is designed for interactive searching by human users using a variety of inputs. It supports simple (Figure 7) and advanced (Figure 8) searches. The award number can be 1) inserted into the simple search, 2) inserted as an Identifier Number in the advanced search, or 3) inserted as an Award ID in the advanced search.

Table 2 shows the number of results (represented by DOIs) that are returned for simple and advanced PAR searches for several award numbers. As articles are identified with DOIs, we use these two terms synonymously. There are many cases where the numbers differ and, in those cases, the simple search typically returns more DOIs than the advanced search. The award number from this small sample with the largest difference between the two searches is 1314642 (bold) with ten DOIs discovered in the simple search and only three in the advanced search.

| Target Award | Search | Results | Target Award | Search | Results |
|---|---|---|---|---|---|
| 2038246 | Simple | 14 | 2028868 | Simple | 12 |
| | Advanced | 11 | | Advanced | 12 |
| 1314642 | Simple | 10 | 2120947 | Simple | 2 |
| | Advanced | 3 | | Advanced | 2 |
| 1805022 | Simple | 36 | 2038246 | Simple | 14 |
| | Advanced | 35 | | Advanced | 11 |

Table 2. Number of results (DOIs) found for award searches using the simple and advanced search interfaces.

Table 3 shows all DOIs associated with award 1314642 in PAR and CHORUS along with the sources where the DOIs were found. The three DOIs found in the PAR advanced search are bold and include the award ID (1314642) in the Award IDs column from PAR data. The 10 DOIs found in the PAR simple search include P in the source in Table 3. It appears that the advanced search for Award Id retrieves only items that include the target award number in the AWARD_IDS field of the metadata, whereas the simple search finds items that are connected to the award but do not include the award Id in the metadata.

| DOI | Year | PAR Award IDs | Source |
|---|---|---|---|
| 10.1038/s41598-023-28166-2 | 2020 | 1840381; 1314642 | P |
| 10.1016/j.hal.2019.101728 | 2021 | 1840381; 1314642; 0911031; 0430724 | P |
| 10.1002/lno.10530 | 2014 | | C&P |
| 10.1021/pr5004664 | 2016 | | C&P |
| 10.1111/eva.12695 | 2017 | | C&P |
| 10.1002/lno.10664 | 2017 | | C&P |
| 10.1038/s42003-021-02626-9 | 2018 | 1840381 | C&P |
| 10.1016/j.hal.2018.08.001 | 2018 | 1314642 | C&P |
| 10.1111/jpy.12386 | 2019 | | C&P |
| 10.1093/toxsci/kfz217 | 2020 | | C&P |
| 10.1016/j.hal.2015.05.010 | 2015 | | C |
| 10.1016/j.ecss.2015.06.023 | 2015 | | C |
| 10.1016/j.hal.2015.07.009 | 2015 | | C |
| 10.1016/j.bbagen.2015.07.010 | 2015 | | C |
| 10.1016/j.ntt.2015.04.093 | 2015 | | C |
| 10.1021/acs.chemrestox.5b00003 | 2015 | | C |
| 10.1016/j.hal.2015.11.003 | 2016 | | C |
| 10.1016/j.neuro.2015.11.012 | 2016 | | C |
| 10.1016/j.taap.2016.02.001 | 2016 | | C |
| 10.1016/j.marpolbul.2016.01.057 | 2016 | | C |
| 10.1093/toxsci/kfx192 | 2017 | | C |
| 10.1016/j.toxicon.2018.06.067 | 2018 | | C |
| 10.1093/eep/dvy005 | 2018 | | C |
| 10.1016/j.hal.2018.06.007 | 2018 | | C |
| 10.1073/pnas.1901080116 | 2019 | | C |
| 10.1093/toxsci/kfab066 | 2021 | | C |

Table 3. DOIs found for award id 1314642 in PAR (P), CHORUS and PAR (C&P), and CHORUS (C). Three DOIs (bold) were discovered in PAR using the advanced search.

Table 3 also includes 24 DOIs associated with the award number 1314642 from the CHORUS All Report from early 2024. DOIs found in PAR and CHORUS are marked with C&P in the source column; those found only in CHORUS are marked with a C. These results clearly indicate that some articles that reference NSF funding in their acknowledgements and, as a result, are indexed in CHORUS, are not included in PAR. PAR and CHORUS have very different data journeys, so it is not surprising that they include different articles. The next section describes an approach to finding these missing articles.

**Comparing CHORUS and PAR Articles**

PAR does not support a web API, but the PAR http interface makes it possible to query for connections between awards and article DOIs using URLs with the form: https://par.nsf.gov/search/term:[awardID]/identifier:[DOI]. If the award number and DOI are connected in PAR, this URL returns information about the DOI as formatted HTML text in the search result. If the award number and DOI are not associated, that text is missing, so the total number of bytes returned is smaller. Thus, the existence of the award-DOI connection could be tested using the response length.

We tested this hypothesis with DOIs found in CHORUS for approximately 50,000 awards, inserting these into the query URL referenced above, and examining resulting response lengths. Figure 9 shows that two distinct groups emerge from these data. The largest group, which includes over 180,000 URLs, has response lengths between 225,500 and 226,000 bytes. The second group is more variable, with lengths distributed over a much larger range – 269,500 to 274,500 bytes. There is a clear gap between these two groups.

This is the behavior we would expect if the awards/DOI combinations in the first group are not included in PAR. They all have essentially the same response – just the framework for the page without any article information. The queries for awards and associated DOIs have much more variable lengths resulting from the HTML with the DOI metadata. The data show that there are substantially more DOIs in the first group – those that are not associated with awards in PAR.

We examined several DOIs manually to test this conclusion. Table 4 shows the results for 11 DOIs associated with award number 2038246 in CHORUS. The response lengths for these DOIs reflect the groups seen in Figure 9, with six lesser than 226,000 and five greater than 269,500. This suggests that five of the 11 DOIs are connected to award 2038246 in PAR (green) and six are not (brown).

| DOI | Publication | Award | Length (bytes) |
|-----|-------------|-------|----------------|
| 10.5194/esd-2021-70 | 2022 | 2021 - 2024 | 225573 |
| 10.5194/acp-2022-372 | 2022 | 2021 - 2024 | 225575 |
| 10.5194/acp-23-687-2023 | 2023 | 2021 - 2024 | 225581 |
| 10.1002/essoar.10509627.1 (preprint) | 2021 | 2021 - 2024 | 225585 |
| 10.5194/egusphere-2023-117 | 2023 | 2021 - 2024 | 225587 |
| 10.1088/2515-7620/acf441 | 2023 | 2021 - 2024 | 225975 |
| 10.1029/2023gl104417 | 2023 | 2021 - 2024 | 271868 |
| 10.1029/2023gl104726 | 2023 | 2021 - 2024 | 271893 |
| 10.1029/2023ef003851 | 2023 | 2021 - 2024 | 272209 |
| 10.1029/2023jd039434 | 2023 | 2021 - 2024 | 272469 |
| 10.5194/esd-13-201-2022 | 2022 | 2021 - 2024 | 272579 |

Table 4. DOIs found in CHORUS for award 2038246 with publication dates, award dates, and response lengths. Six of these DOIs are in CHORUS, but not in PAR (brown). Five are in CHORUS and in PAR (green).

The timing of these publications with respect to the award period is important because NSF researchers are only expected to add items to PAR during the award period. The articles in Table 3 that are missing from PAR were all published during the award period for award 2038246. Manual checking confirmed that these six papers directly acknowledge NSF award 2038246 in their text (Table 5), consistent with the observation that they are in CHORUS with that award number.

| DOI | Acknowledgement (manual search) |
|---|---|
| 10.5194/esd-2021-70 | Support for Y. Zhang and D. G. MacMartin was provided by the National Science Foundation through agreement CBET-1818759 and CBET-2038246. |

| 10.5194/acp-2022-372 | National Science Foundation through agreement CBET-1818759 for DV and DGM |
|---|---|
| 10.5194/acp-23-687-2023 | National Science Foundation through agreement CBET-2038246 for Douglas G. MacMartin |
| 10.1002/essoar.10509627.1 | National Science Foundation through agreement CBET-1818759 and CBET-2038246. |
| 10.5194/egusphere-2023-117 | National Science Foundation (grant no. CBET-2038246) |
| 10.1088/2515-7620/acf441 | National Science Foundation through agreement CBET-1931641, as well as CBET-2038246 for DGM and BK |

Table 5. NSF acknowledgements from papers in CHORUS that are missing from PAR.

We tested 51,602 of 97,485 awards (53%) from CHORUS and found DOI matches in PAR for 32,543 of the tested awards (63%), leaving 19,059 awards with associated articles in CHORUS that have no published results in PAR. The tested awards are associated with 307,549 DOIs in CHORUS and 127,218 (41%) of these were found in PAR. The numbers of DOIs found in PAR are shown in green in Figure 10 along with DOIs found in CHORUS and not in PAR as blue.

It is important to explore temporal variations of these data as their data journeys are complicated and evolve over time. Figure 11 shows relative number of DOIs in three groups: found in CHORUS and PAR (green), found in CHORUS only (blue), and not included in this test (grey) as a function of Award Effective Year. The fall-off on the recent end of the data reflects the delay in publishing results or including publications in PAR for recent awards.

These data show a sharp increase in the portion of DOIs that have been included in PAR for awards effective during 2017 or later. This increase is more evident in the stacked bar plot shown in Figure 12. Between 2000 and 2016, an average 25% of the DOIs in CHORUS were also in PAR. This increased to 64% between 2017 and 2023.

## Awards

### Comparing CHORUS and PAR

For the purposes of this study, data downloaded from the NSF Award Search database are taken to represent not only the total corpus of NSF grants awarded in a given period but also the "truth" of those awards. In the NSF Award Search database, award numbers are stored in a field called "Award – AwardID". They are stored as strings of seven digits only, with directorate-level information stored in separate fields.

In the CHORUS All report, award numbers are stored in the "Grant ID" field. As mentioned above, this field captures award numbers for all funders – the values it stores therefore demonstrate a wide variety of syntax. After applying the regular expressions methodology mentioned above, a character count analysis was run over CHORUS Grant ID values believed to be NSF award numbers. This analysis revealed that the most common syntax of NSF Award numbers in CHORS data is "NSF:ABC-1234567", where "NSF:" is prepended by CHORUS and the remainder is the value provided by the authors to their respective publishers as described in the CHORUS data journeys above. The middle three-letter abbreviation generally reflects the NSF division or directorate which made the grant. While this is the most common way that grants are referenced in Crossref metadata (and CHORUS), variations abound, with some authors opting to spell things out ("NSF: National Science Foundation:CHE-1205646") and others forgoing numeric reference altogether ("NSF:ACLS:Dissertation Completion Fellowship").

In PAR, award numbers are stored in a field called "Award_ID". The values stored in this field refer to NSF awards only, utilizing the proper seven-digit syntax 100% of the time. This consistency in formatting supports a relatively straightforward analysis of the proportion of PAR records that contain NSF award numbers. Of the 186,526 total records in our PAR sample, 134,807 (72%) contain NSF award numbers. By contrast, our analysis suggests that of the 412,441 total DOIs in the CHORUS All report, 280,432 (68%) contain NSF award numbers. These numbers reflect comparisons run across all PAR records through June 2022, all records contained in the February 2024 CHORUS All report, and all standard, continuing and fellowship awards granted by NSF between 2004 and 2021.

Comparison of the Award Database and PAR

The numbers above afford some insight into the completeness of PAR and CHORUS metadata, focusing specifically on the completeness of funding information across each source. By shifting the center of analysis from PAR and CHORUS to the NSF Award Search database itself, it is also possible to gain a view into the proportion of NSF awards that are referenced by NSF-funded research outputs. To do this, we isolated the NSF award numbers for all standard, continuing and fellowship grants awarded between 2004 and 2021, and used vertical lookups in Excel to identify which awards were referenced in PAR and CHORUS records. We analyzed these first as a whole and found that of the 211,012 such awards granted by NSF during that time frame, 30,297 (14%) are referenced in PAR records while 75,594 (36%) are referenced in CHORUS records, leaving 118,528 (56%) unreferenced across either source. The percentages reported here add up to 106 because there is an overlap of 6%; of the 92,484 awards referenced in either PAR or CHORUS records, 13,407 (6%) are in fact referenced across both (Figure 13).

The data journeys described above provide some insight into the differences between the groups in Figure 13. For an item to be included in PAR or in CHORUS, it must be peer-reviewed and published. If a published work acknowledges funding with a funder name or identifier in Crossref metadata, it is included in CHORUS. So, the difference between PAR Only (yellow) and CHORUS & PAR (green) or CHORUS Only (blue) is related to funder acknowledgement in Crossref. Items in PAR Only are peer-reviewed, published, and submitted to PAR, while items in the CHORUS groups are peer-reviewed, published, and acknowledged in Crossref metadata whether or not they are submitted to PAR. Items in the largest group (grey), are not submitted to PAR and, if they are published, do not acknowledge NSF.

**History**

As is illustrated by Figure 14, these numbers gain additional depth when considered in the context of time. This Figure shows pie charts like Figure 13 for years following award effective dates of 2014-2018. The top row starts during 2014 and extends until 2021 while the bottom row starts in 2018 and extends until 2021.

The percentages in these charts are cumulative. For example, 11,758 standard, continuing and fellowship awards granted by NSF went into effect in 2014 (top row). As time passes and references to these awards occur in CHORUS or PAR or both, the percentages of unreferenced awards (No Reference) decreases while the other categories grow. After 4 years, 51% of these 2014 awards are not referenced, 43% are in CHORUS only, 6% are in CHORUS and PAR, and 1% are only in PAR.

These plots clearly show an evolution of references to NSF awards over time. Awards that became effective during 2014 (top row) are referenced primarily only in CHORUS, even after seven years. By 2021 (dashed line in Figure 14), 41% remain unreferenced, 52% are only in CHORUS, 7% are in CHORUS and PAR, and 1% are only in PAR. Awards that became effective in 2018 (bottom row), get referenced much more quickly and the distribution of references changes significantly. By 2021only 36% remain unreferenced, 7% are only in CHORUS, 40% are in CHORUS and PAR, and 17% are only in PAR. This trend towards more awards referenced more quickly develops over time with the percentage of unreferenced awards after three years dropping from 59% for 2014 awards to 36% for 2018 awards. Three years is the typical length of an NSF award and PAR reporting is only required during the active award period.

As described above and illustrated in Figure 13, references to awards in PAR and CHORUS reflect different data journeys. The increase in award coverage in PAR, indicated by the increase in the size of the yellow slices for award effective dates 2017 and 2018 in Figure 14, reflects an increase in the number of NSF awards that supported peer reviewed and published papers, but are not acknowledged in those papers in a way that is reflected in Crossref metadata. If acknowledgements were there, they would be referenced in CHORUS and be included in the blue or green slices.

Figure 15 shows the evolving distribution of NSF award references during 2021 for awards with effective dates between 2014 and 2018, the five most recent years on the right edge of Figure 14. The oldest awards, effective during 2014, were overwhelmingly referenced in CHORUS rather than PAR, while the most recent awards, effective during 2018, are mostly referenced in PAR or in CHORUS and PAR. Roughly one third of the recent awards remain unreferenced.

## Conclusions

The NSF Award Search and Public Access Repository (PAR) are important public resources for identifying NSF awards and related research results. The recent emergence of linked repositories of identifiers and metadata for articles, datasets, researchers, research organizations, and funders (termed the "global research infrastructure") provides another view which is populated through a complex and evolving set of data journeys largely outside of the purview of NSF. CHORUS combines several repositories from the global research infrastructure, as well as the connections between them, to create publicly available datasets of published articles and datasets supported by NSF and other funders. The INFORMATE Project compared these three sources to understand and characterize the contributions they make to finding NSF funded results.

The CHORUS dataset includes significantly more awards and more related papers than PAR and, as a more complete source, it could be used to find published research results that are not included in PAR. We searched PAR for award/article pairs found in CHORUS and identified over 19,000 awards with related articles that were not included in PAR and over 180,000 published articles that acknowledged NSF funding that were not included in PAR.

Temporal analysis shows that many of these missing articles are associated with awards with effective dates between 2014 and 2016. Awards with effective dates between 2017 and 2018 are more likely to be included in both CHORUS and PAR although between 5 and 10% of the awards during this time are not referenced in PAR and roughly one third of the recent awards are not referenced in CHORUS or PAR. Recent developments in full text processing, artificial intelligence, and natural language processing, such as those described by Dumanis et al., provide a promising approach to finding contributions from these awards and understanding the nature of award references that are in PAR but not in CHORUS (those that are, by extension, not acknowledged clearly in the literature).(20)

CHORUS and PAR have very different data journeys, each presenting different obstacles. In the CHORUS case, correct funder names (without acronyms) and funder Ids are required in Crossref for articles to be discovered in funder searches and correct award numbers are required for association between an article and an award. Authors must ensure that these metadata make it to journal publishers and are correctly migrated to Crossref.

In the PAR case, researchers must make a conscious decision to add resources to PAR and enter article metadata correctly and completely for each resource connected to an award, even if it is published after the award ends. This process has been greatly improved recently with the capability for researchers to enter a DOI and let the system retrieve the complete metadata. Even prior to this development, there have been increases in the completeness of PAR (Figure 15), potentially reflecting increased community awareness and adoption.

Another difference between these data journeys is the level of automation. The CHORUS journey involves more underlying systems and connections between them, but it is consistent, verifiable and reproducible. The PAR journey involves more human activity and is less consistent across resource collections, adding uncertainty to interpreting changes.

These comparisons demonstrate that even where complex data journeys exist, the global research infrastructure can provide data that can significantly increase the breadth of identified NSF funded work and provide more complete estimates of the impact of NSF funding.

## Acknowledgements

This research was funded by U.S. National Science Foundation award 2334426. The authors wish to thank Howard Ratner and Tara Packer of CHORUS for their input and feedback throughout the course of this project.

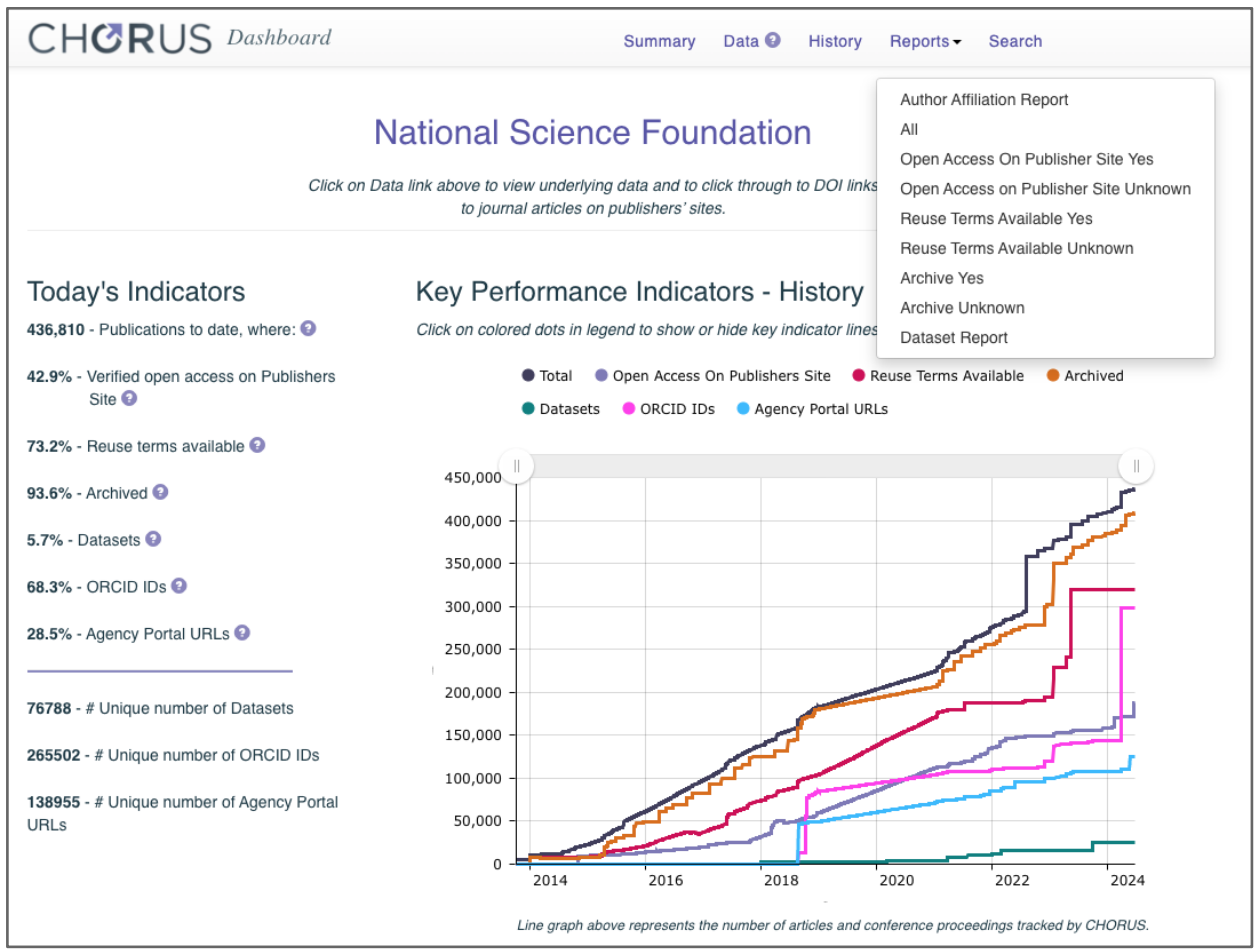

Figure 1. CHORUS dashboard for the US National Science Foundation with time histories of monitored parameters and report selector list. This work focused on the All, Author, and Dataset reports.

Figure 2. CHORUS reports considered here and metadata they include.

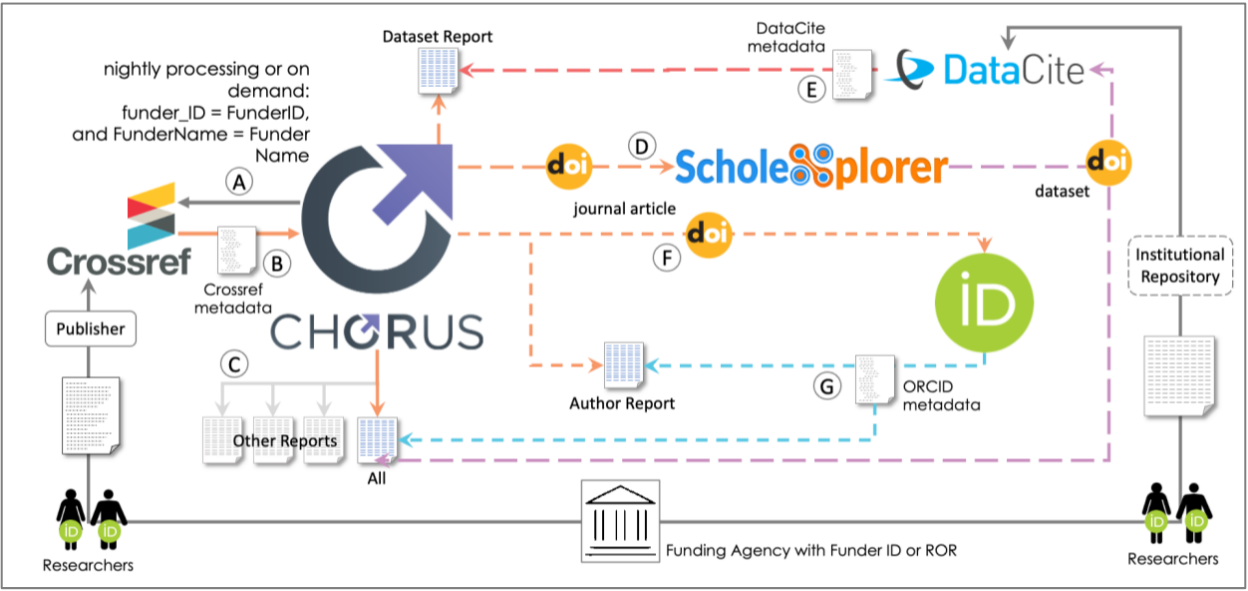

Figure 3. The CHORUS data journey

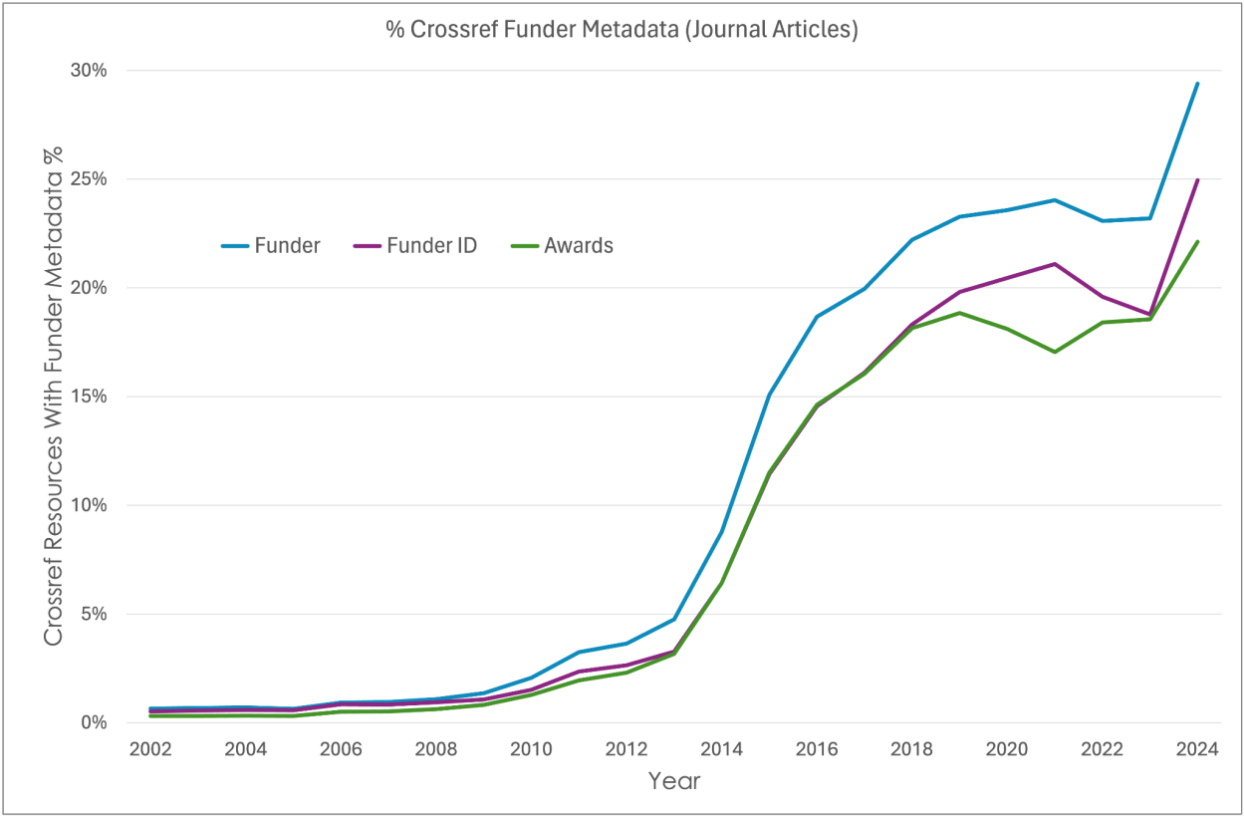

Figure 4. History of Crossref funder metadata completeness for journal articles

| | A | B | N |
|---|---|---|---|
| 1 | result - osti_id | result - author - author_lname | result - title |
| 2 | 10021311 | Reginato | The first complete plastid genomes of Melastomataceae are highly structurally conserved |
| 3 | 10021311 | Neubig | The first complete plastid genomes of Melastomataceae are highly structurally conserved |
| 4 | 10021311 | Majure | The first complete plastid genomes of Melastomataceae are highly structurally conserved |
| 5 | 10021311 | Michelangeli | The first complete plastid genomes of Melastomataceae are highly structurally conserved |
| 6 | 10021525 | Roux | Towards quantitative viromics for both double-stranded and single-stranded DNA viruses |
| 7 | 10021525 | Solonenko | Towards quantitative viromics for both double-stranded and single-stranded DNA viruses |
| 8 | 10021525 | Dang | Towards quantitative viromics for both double-stranded and single-stranded DNA viruses |
| 9 | 10021525 | Poulos | Towards quantitative viromics for both double-stranded and single-stranded DNA viruses |
| 10 | 10021525 | Schwenck | Towards quantitative viromics for both double-stranded and single-stranded DNA viruses |
| 11 | 10021525 | Goldsmith | Towards quantitative viromics for both double-stranded and single-stranded DNA viruses |
| 12 | 10021525 | Coleman | Towards quantitative viromics for both double-stranded and single-stranded DNA viruses |
| 13 | 10021525 | Breitbart | Towards quantitative viromics for both double-stranded and single-stranded DNA viruses |
| 14 | 10021525 | Sullivan | Towards quantitative viromics for both double-stranded and single-stranded DNA viruses |

Figure 5. PAR data demonstrating the one-to many relationship between publications and authors. Each result - osti_id value is unique to a publication; rows are repeated to capture repeating elements such as author names.

| result - osti_id | result - doi |
|---|---|
| 10320686 | 10.1525/elementa.2020.20.00055 |
| 10282470 | -10.1140/epjc/s10052-020-08801-2 |
| 10217486 | : 10.1017/jfm.2020.708 |
| 10314320 | .2021.107607 0888-3270 |
| 10140225 | .doi.org/10.1080/87565641.2019.1688328 |
| 10314677 | ‚Ää‚Ää10.1016/j.parco.2021.102793 |
| 10232187 | //doi.org/10.1063/5.0045004 |
| 10297838 | /j.jfranklin.2021.04.001 |
| 10042448 | ¬†10.1109/ISIT.2017.8006821 |
| 10143109 | OE.26.025534 |
| 10230356 | Remote Power Side-Channel Attacks on BNN Accelerators in FPGAs |
| 10041476 | RG.2.2.17468.74883 |
| 10237549 | s00222-020-00962-x |
| 10122019 | tp://dx.doi.org/10.5951/mathteaceduc.8.1.0076 |

Figure 6. Representative content and formatting errors in NSF PAR DOI values.

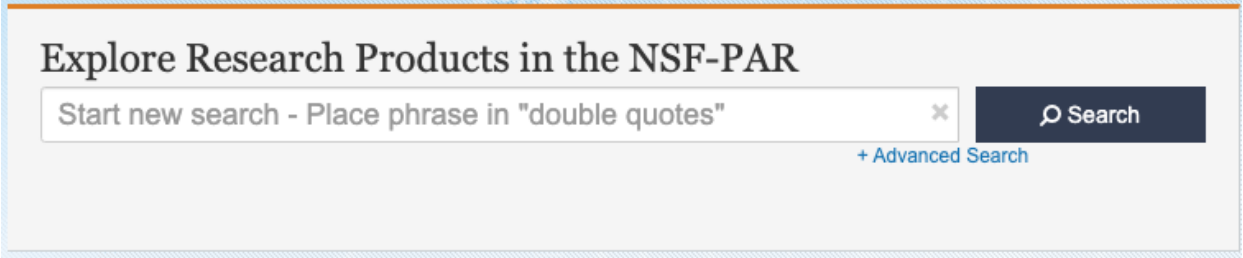

Figure 7. Simple CHORUS search interface.

Figure 8. Advanced PAR search interface. Note specific input field for Award ID.

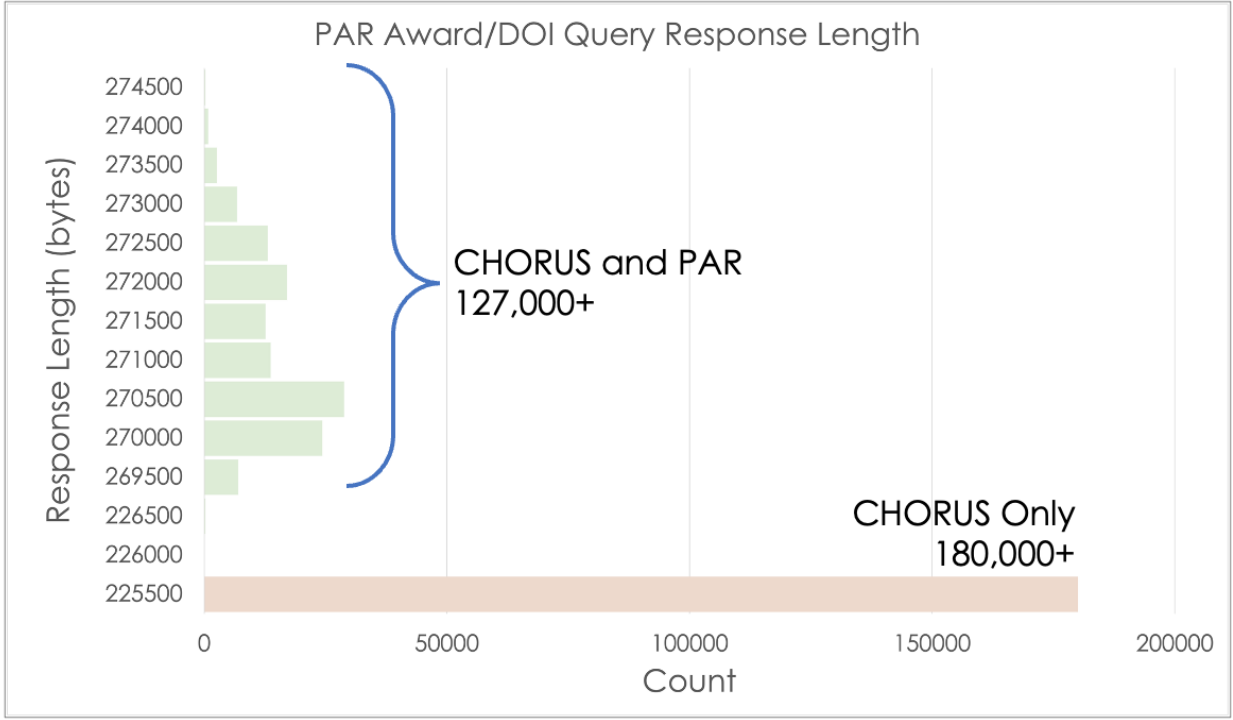

Figure 9. Response lengths for over 300,000 PAR queries for awards and DOIs.

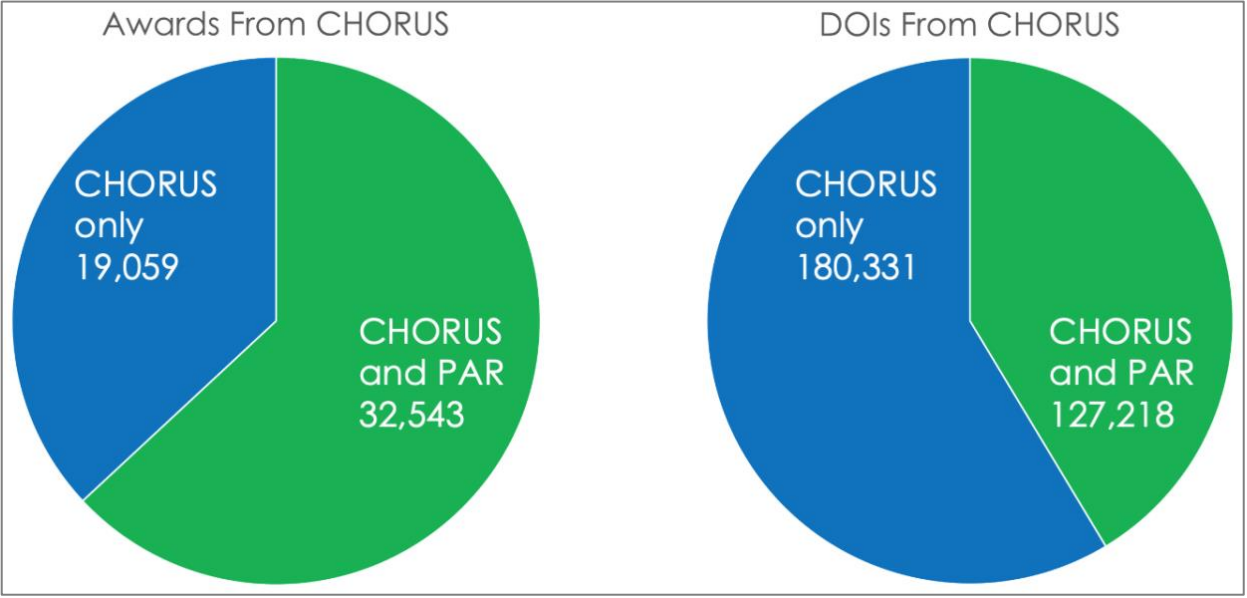

Figure 10. Numbers of CHORUS Awards and DOIS found in PAR for a test set of 51,602 awards.

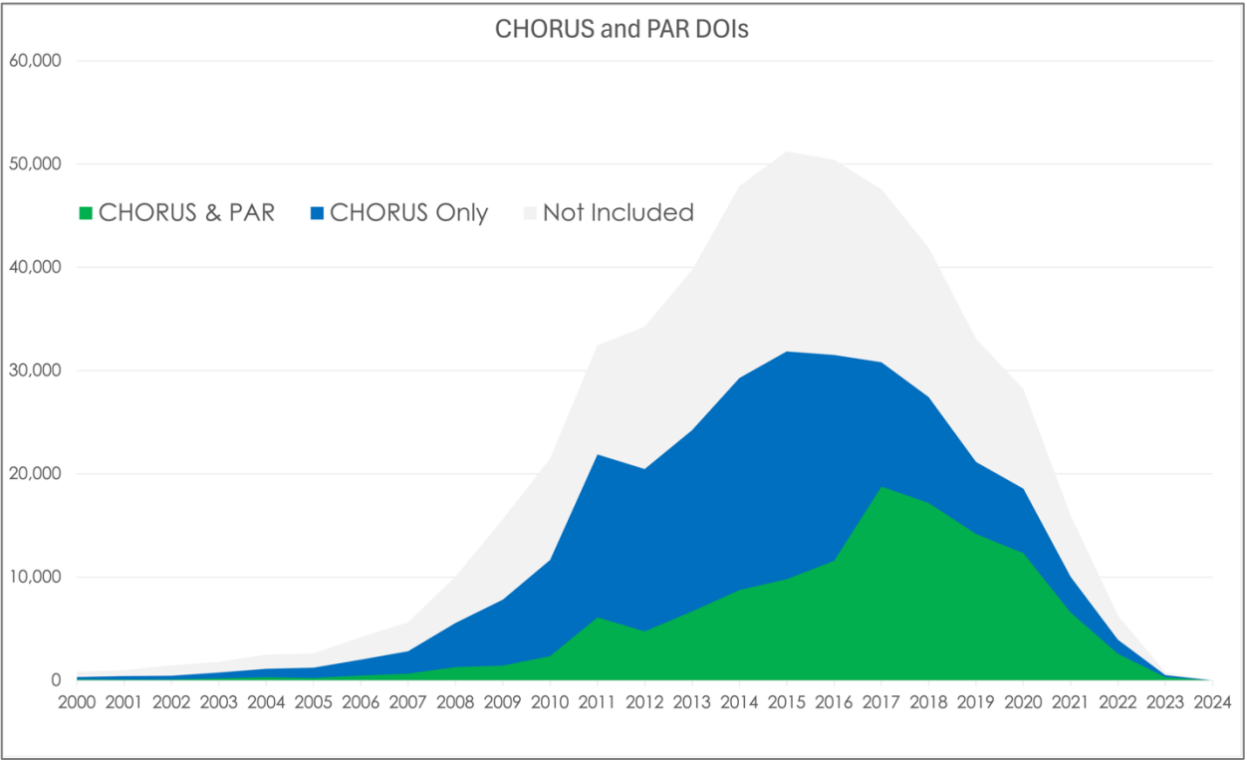

Figure 11. Stacked timeseries of DOIs associated with NSF awards from CHORUS and PAR, CHORUS only, and Not Included by Award Effective Date.

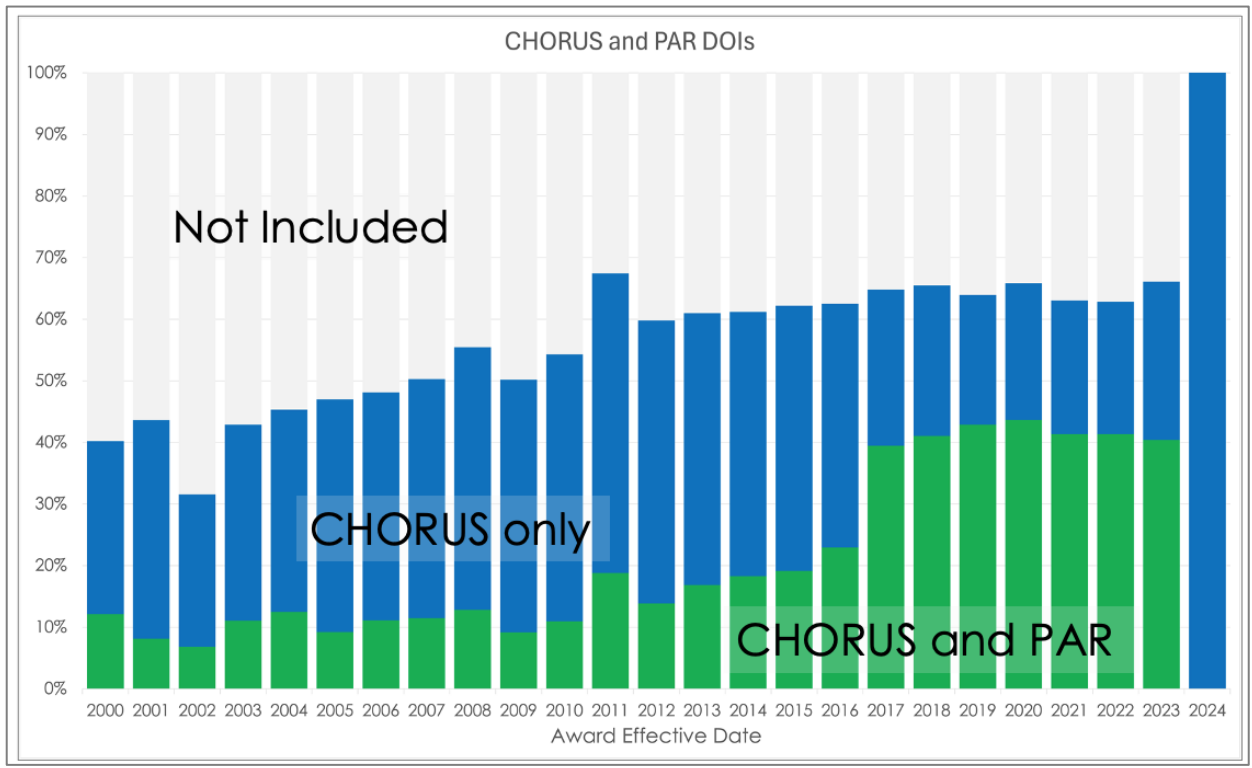

Figure 12. Percentage of DOIs per year found in CHORUS (blue), CHORUS and PAR (green) and not included in the test (Grey). Note the jump in PAR coverage during 2017.

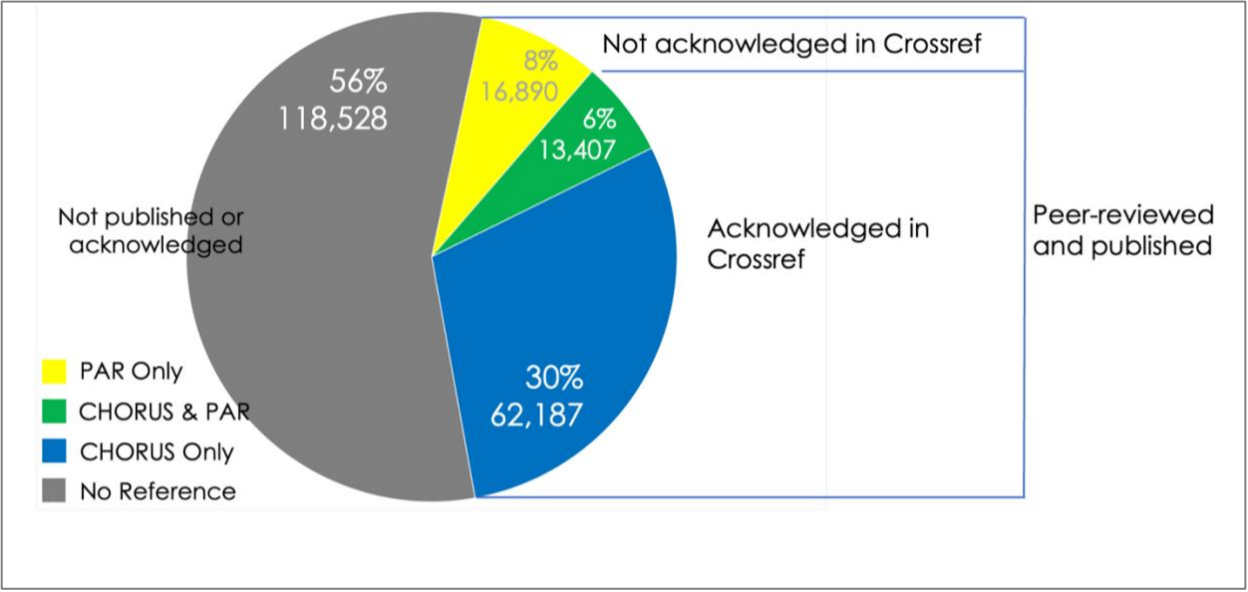

Figure 13. Distribution of references to NSF award numbers across PAR and CHORUS (Standard, Continuing and Fellowship awards granted 2004-2021) with data journeys.

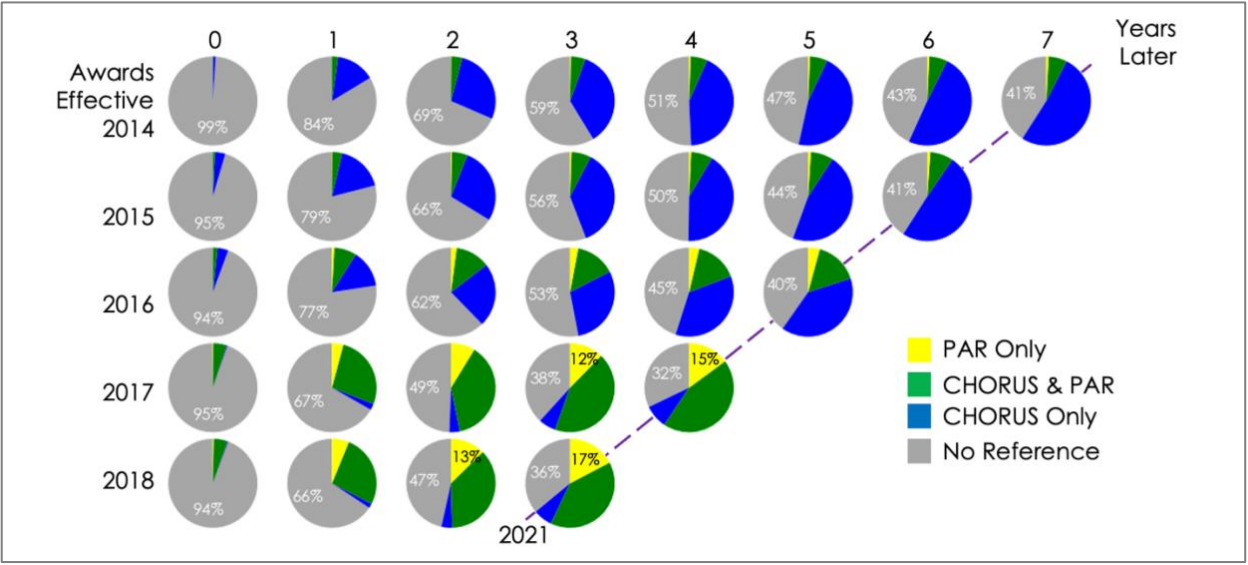

Figure 14. NSF awards referenced in PAR and CHORUS by year. The Y-axis is Award Effective Date, and the X-axis is Years After Award Effective Date. Real years are on a slanted axis shown for 2021.  The percentage of NSF awards referenced in CHORUS (blue), CHORUS and PAR (green), PAR (yellow) and not referenced (gray) are shown. Percentage values are shown for No Reference every year and for PAR if the percentage is greater than 10.

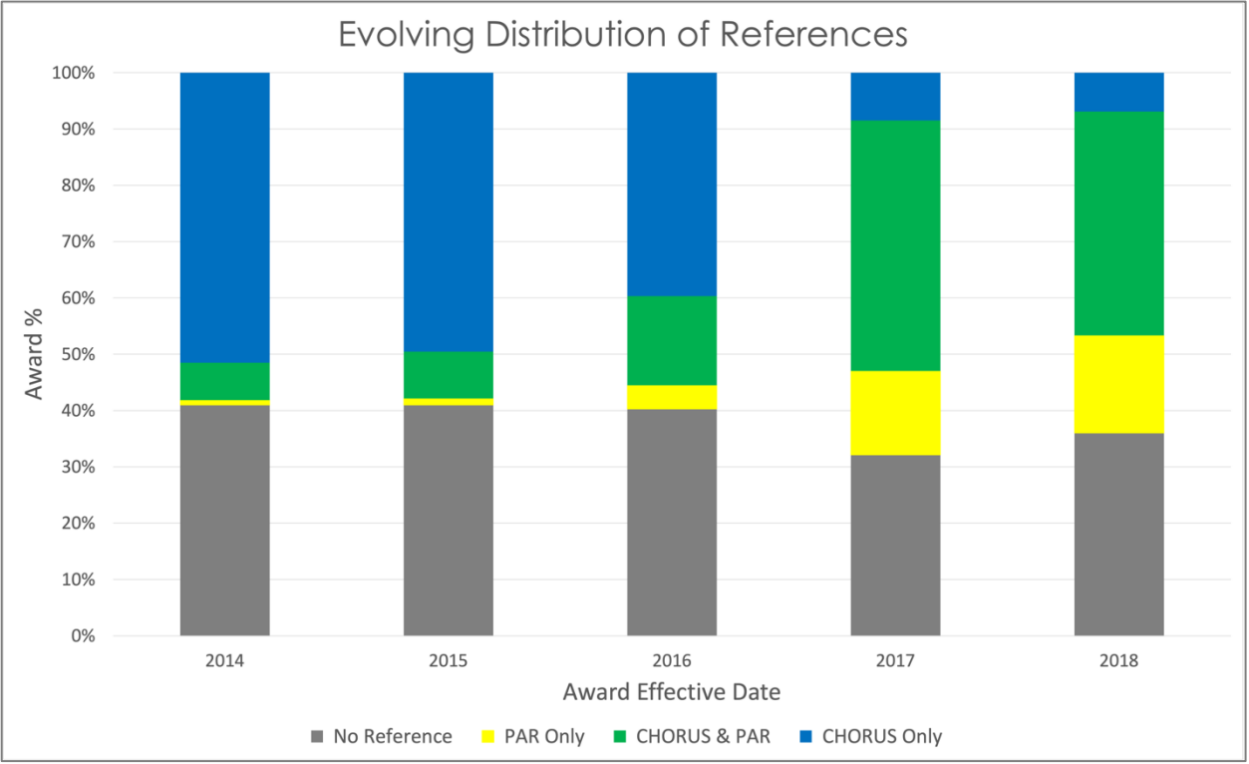

Figure 15. The distribution of NSF award references during 2021 (slanted dashed line in (Figure 14)) for awards with effective dates between 2014 and 2018.